\documentclass[cits]{PoS}
\pdfoutput=1
\title{Effects of the low lying Dirac modes on the spectrum of ground
state mesons}

\ShortTitle{Effects of the low lying Dirac modes on the spectrum of ground
state mesons}

\author{C.B. Lang\\
        Institut f\"ur Physik, FB Theoretische Physik, Universit\"at
Graz, A--8010 Graz, Austria\\
       E-mail: \email{christian.lang@uni-graz.at}}

\author{\speaker{Mario Schr\"ock}\\
        Institut f\"ur Physik, FB Theoretische Physik, Universit\"at
Graz, A--8010 Graz, Austria\\
       E-mail: \email{mario.schroeck@uni-graz.at}}

\abstract{The lowest eigenmodes of the Dirac operator are related to the 
dynamical breaking of the chiral symmetry in Quantum Chromodynamics (QCD).
In our work we construct quark propagators which exclude a varying number of the lowest
Dirac eigenmodes and study the influence thereof on meson correlators and the meson spectrum.
That procedure partially restores the chiral symmetry 
(in the valence sector) and we observe degeneracies
in the spectrum while confinement seems not to be affected.
}

\FullConference{XXIX International Symposium on Lattice Field Theory \\
		 July 10--16, 2011\\
		 Squaw Valley, Lake Tahoe, California}

\usepackage[tight]{units}
\usepackage{amsmath}
\usepackage{wrapfig}

\newcommand{\fig}[1]{Fig.~{\ref{#1}}}

\newcommand{\FD}{\;.}

\newcommand{\RDH}[1]{{\mathrm{red5}(#1)}}
\newcommand{\LM}[1]{{\mathrm{lm}(#1)}}
\newcommand{\LMH}[1]{{\mathrm{lm5}(#1)}}
\newcommand{\SU}[1]{\mathrm{SU}(#1)}
\newcommand{\U}[1]{\mathrm{U}(#1)}

\newcommand{\ubar}{\overline{u}}
\newcommand{\ket}[1]{\left|{#1}\right>}
\newcommand{\bra}[1]{\left<{#1}\right|}
\newcommand{\be}{\begin{equation}}
\newcommand{\ee}{\end{equation}}
\newcommand{\gaf}{\gamma_5}

\begin{document}

\section{Motivation}
The Banks-Casher relation \cite{Banks:1979yr} associates the density of the 
smallest non-zero eigenvalues to the chiral condensate. The importance of the
lowest Dirac eigenmodes in QCD has been investigated within several studies: in
\cite{DeGrand:2000gq,DeGrand:2001tm,DeGrand:2003sf} it was shown that the
pseudoscalar and axial vector correlators can be approximated in the medium to
large temporal distance region when taking only the lowest Dirac eigenmodes as input for the
quark propagators into account. In
\cite{DeGrand:2003sf,DeGrand:2004qw,DeGrand:2004wh,Giusti:2004yp} this was used
to improve the signal of hadron correlators by averaging the low mode
contribution to the quark propagator over all possible positions of the sinks on
the lattice. The authors of \cite{DeGrand:2003sf,DeGrand:2004qw} 
studied the influence of low mode removal in the quenched approximation using
the overlap operator \cite{Neuberger:1997fp,Neuberger:1998wv}.

In our study \cite{Lang:2011qy} we construct meson correlators out of \emph{reduced} quark propagators
which exclude a varying number of the lowest Dirac eigenmodes.
Using these propagators we compute meson masses and we study
the possible recovery of the degeneracies of (would
be) chiral partners in the spectrum. 
Moreover this gives
us insight on the role of the lowest Dirac eigenmodes for confinement and
thus on the relation of chiral symmetry and confinement.
Our modification is  applied  only to the valence quark sector; the
configurations have been obtained with 
two dynamical light quark flavors without truncation.

The two lightest quark flavors in QCD are much lighter than the typical QCD scale.
Neglecting the mass of these
two light quarks, the Lagrangian is invariant under the symmetry group
\begin{equation}
	\SU{2}_L\times\SU{2}_R\times\U{1}_V\times\U{1}_A\FD
\end{equation}
The single flavor symmetry $\U{1}_V$ conserves the baryon number density. 
The axial vector part of the chiral symmetry $\SU{2}_L\times\SU{2}_R$ mixes states with opposite parity;
the nondegenerate masses of parity partners in nature indicate the dynamical breaking of the axial symmetry whereas
the isospin symmetry is (approximately) preserved in the vacuum.
The single flavor axial symmetry $\U{1}_A$ mixes the currents of the
same isospin but opposite parity. The latter is not only broken spontaneously
but also explicitly by the anomaly.

\begin{wraptable}{r}{0.45\textwidth}
\vspace{-0.5cm}
\centering
\begin{tabular}{|r|c|c|c|}
	\hline
	   \#  & meson & $J^{PC}$ & interpolator \\\hline\hline
	1&$\pi$ & $0^{-+}$ & $\ubar\gamma_5 d$\\\hline
	2&$\pi$ & $0^{-+}$ & $\ubar\gamma_4\gamma_5 d$\\\hline
	3&$\rho$ & $1^{--}$ & $\ubar\gamma_i d$\\\hline
	4&$\rho$ & $1^{--}$ &  $\ubar\gamma_4\gamma_i d$\\\hline
	5&$a_0$ & $0^{++}$ & $\ubar d$ \\\hline
	6&$a_1$ & $1^{++}$ & $\ubar\gamma_i\gamma_5 d$\\\hline
\end{tabular}\caption{The mesons and the corresponding interpolating fields studied in this work.}
\vspace{-0.5cm}
\end{wraptable}\label{tab:mesontab}
We will investigate the influence of low mode reduction on the pseudoscalar, vector, scalar and axial vector 
currents with the interpolating fields given in Table \ref{tab:mesontab}. 
In a chirally symmetric world, the currents number three and six of the table
would get mixed via the isospin axial transformation and
the currents one and five would get mixed via the single flavor axial transformation $\U{1}_A$.

We did not include isoscalars in our work (these need disconnected terms, notoriously difficult
to compute), otherwise we could additionally study the possible
recovery of degeneracies in the spectrum between the $h_1$ and current four, 
between $f_0$ and current one as well as between the $\eta$ and current five, 
which all would pairwise mix via the axial vector transformation of the chiral symmetry.
For a detailed discussion of the mixing of different currents under the chiral symmetry see \cite{Glozman:2007ek}.

\section{Method}
Most lattice Dirac operators $D$ (an exception is the overlap operator \cite{Neuberger:1997fp,Neuberger:1998wv})
are nonnormal operators, thus one has to distinguish 
between left eigenvectors $\bra{L_i}$ and right eigenvectors $\ket{R_i}$ to a given (complex valued)
eigenvalue $\lambda_i$. Then the spectral representation of $D$ reads
\be\label{Dspectral}
	D = \sum_{i=1}^N \,\lambda_i\,\ket{R_i}\bra{L_i}.
\ee
The hermitian Dirac operator $D_5\equiv \gamma_5D$
is normal and has real eigenvalues $\mu_i$. Alternatively $D$ can be written in terms of the spectral representation of $D_5$ in the following way:
\be\label{gafDspectral}
	D = \sum_{i=1}^N \,\mu_i\,\gamma_5\,\ket{v_i}\bra{v_i}.
\ee

We want to calculate the meson spectrum using reduced quark propagators which leave out a part of the low lying Dirac
spectrum.
Therefore we split the quark propagator $S=D^{-1}$ into a low mode part (lm) and a reduced part (red), e.g., using the eigenvalues and eigenvectors of $D_5$,
\be
	S = \sum_{i\leq k}\, \mu_i^{-1} \,\ket{v_i}\bra{v_i}\,\gaf 
	  + \sum_{i>k}\,\mu_i^{-1}\,\ket{v_i}\bra{v_i}\,\gaf
	  \equiv S_\LMH{k} + S_\RDH{k}.
\ee 
Hence we can obtain the reduced part of the propagator by subtracting the low mode part from the full propagator
\be 
	S_\RDH{k} = S - S_\LMH{k}.
\ee
Note that the low mode part has to operate on the same sources as $S$ in order to obtain
the correct reduced propagators.
The corresponding separation using the left and right eigenvectors and the eigenvalues of $D$ 
can be done in an analogous way.

\section{The setup}

For our calculation we used 161 gauge field configurations \cite{Gattringer:2008vj,
Engel:2010my} of  lattice size $16^3\times 32$. The lattice spacing is
$a=\unit[0.144(1)]{fm}$ and thus the spatial lattice size $L=\unit[2.3]{fm}$.
The configurations are obtained with two light degenerate dynamical quark flavors with
an AWI-mass of $m=\unit[15.3(3)]{MeV}$ and a corresponding pion mass of $m_\pi=\unit[322(5)]{MeV}$.
For the dynamical quarks of the
configurations as well as for the valence quarks of our study the so-called Chirally
Improved Dirac operator \cite{Gattringer:2000js, Gattringer:2000qu} has been
used. This operator is an approximate solution to the Ginsparg--Wilson equation
and therefore exhibits better chiral properties than the Wilson Dirac
operator while being less expensive  by an order of magnitude in comparison to the  chirally exact overlap operator.

\begin{figure}[h]
\centering
\includegraphics[width=0.5\columnwidth]{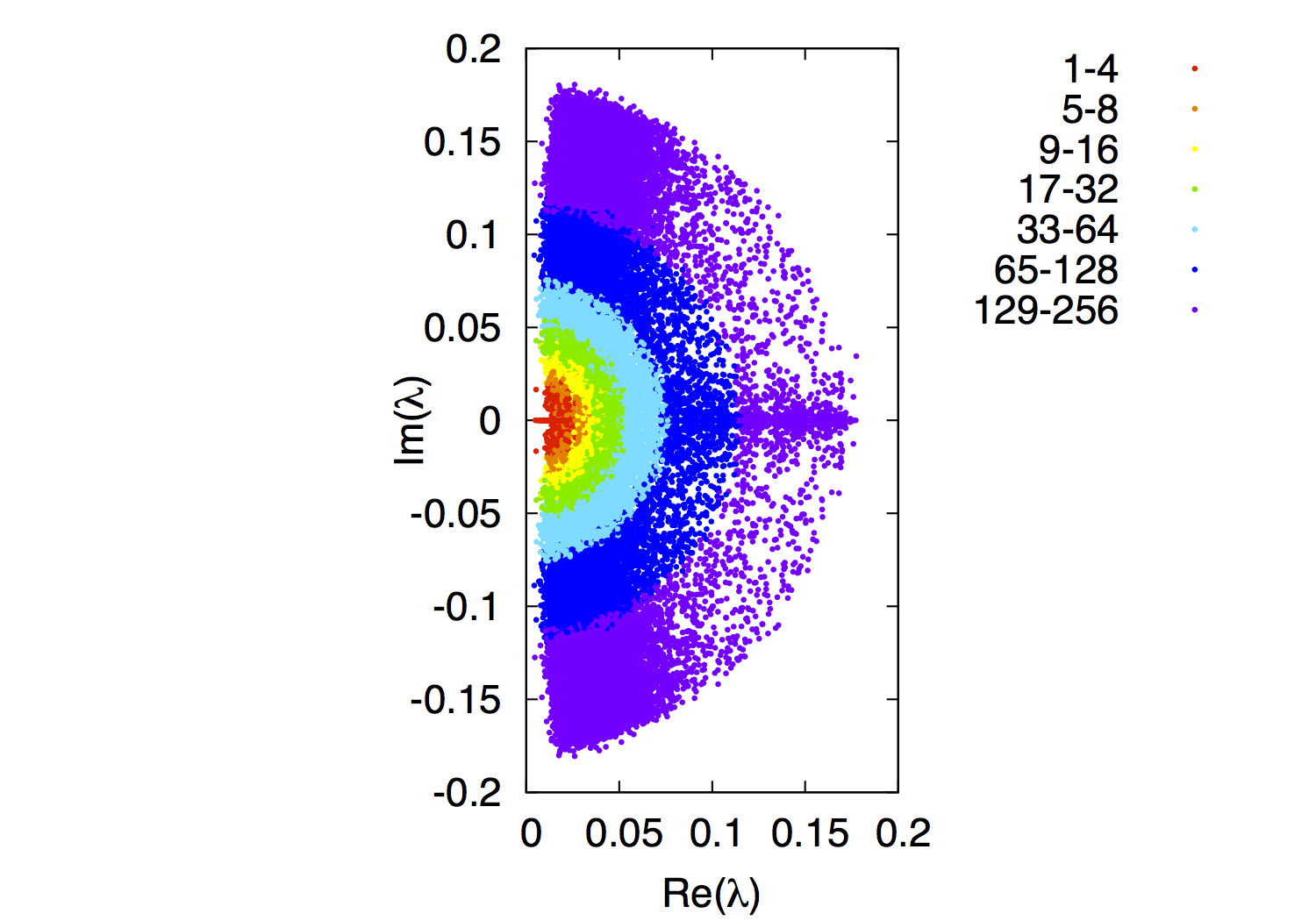}\hfill
\includegraphics[width=0.5\columnwidth]{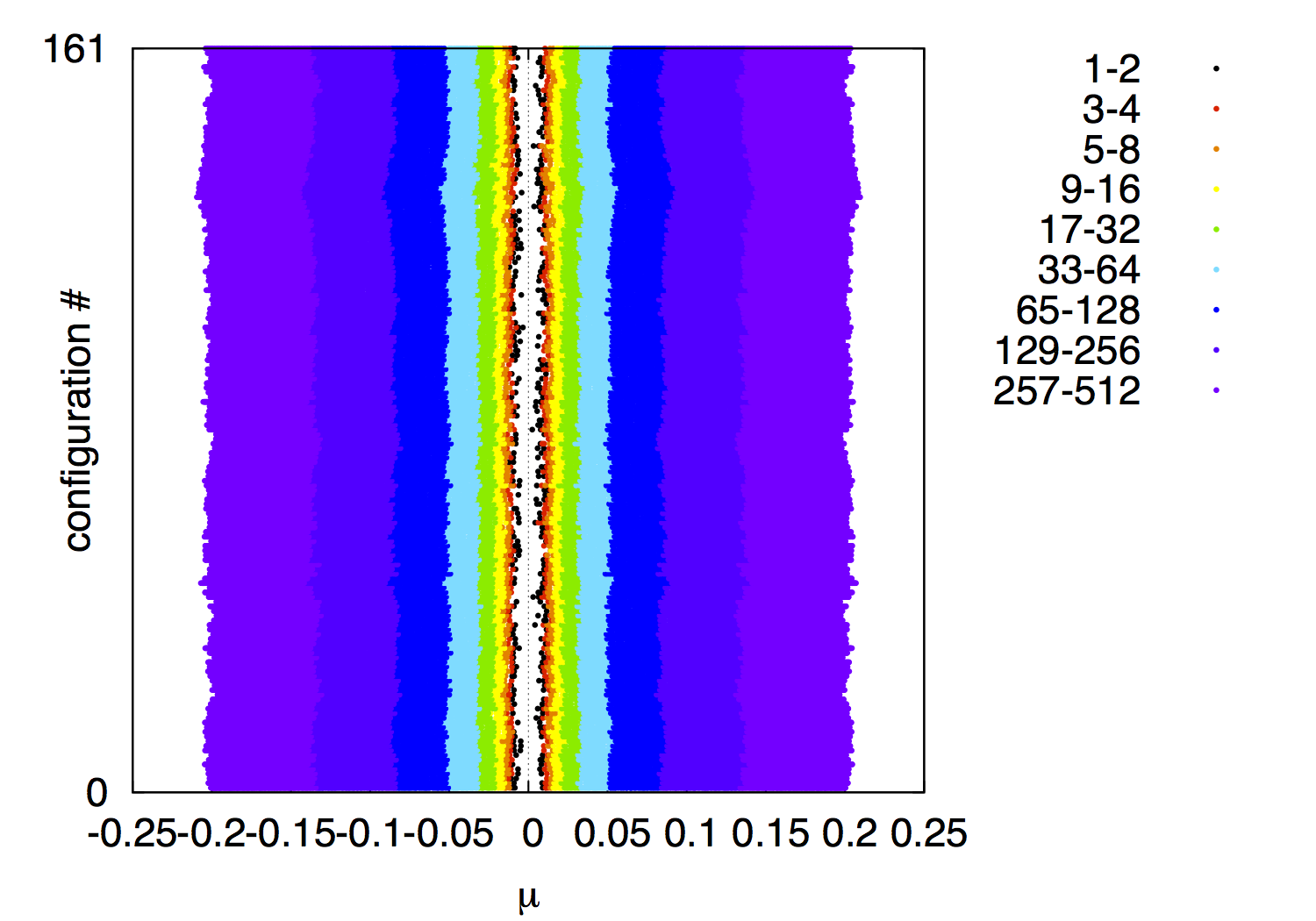} 
\caption{The lowest 256 eigenvalues of $D$ (left) and the lowest 512 eigenvalues of $D_5$ (right) for all 161 configurations.}\label{eigvalues}
\end{figure}

On these configurations we computed the lowest 256 eigenvalues of $D$ and the lowest $512$ eigenvalues of $D_5$,
see \fig{eigvalues}, using ARPACK which is an
implementation of the Arnoldi method to calculate a part of the spectrum of
arbitrary matrices \cite{LeSoYa98}.

\begin{figure}[h]
\centering
\includegraphics[width=0.5\columnwidth]{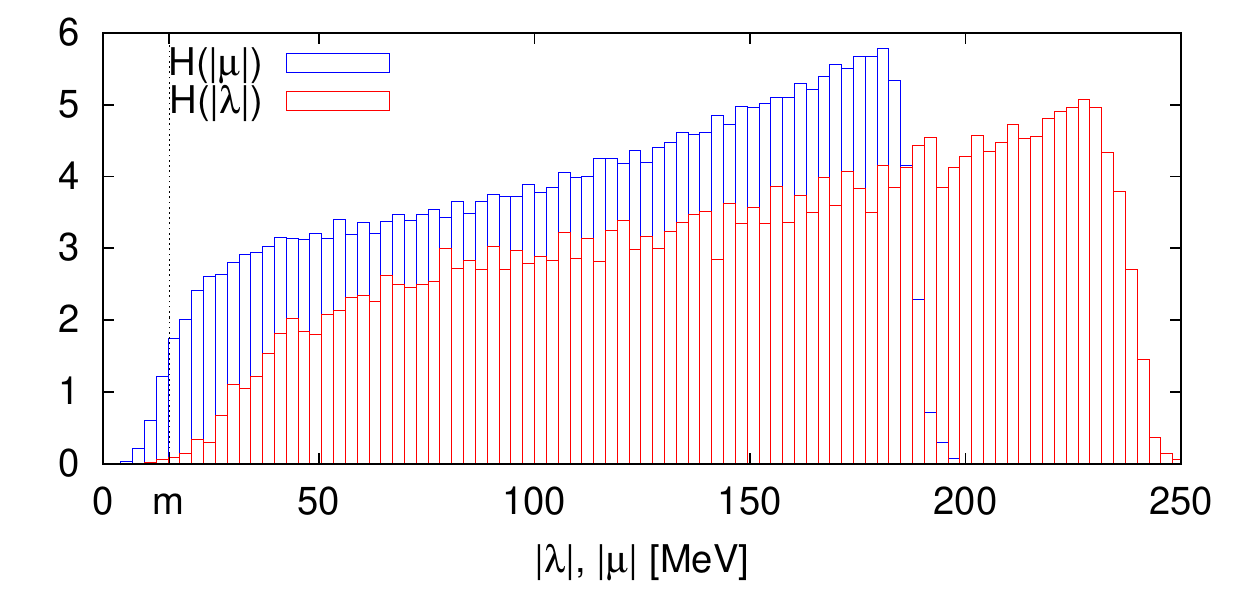}\hfill
\includegraphics[width=0.5\columnwidth]{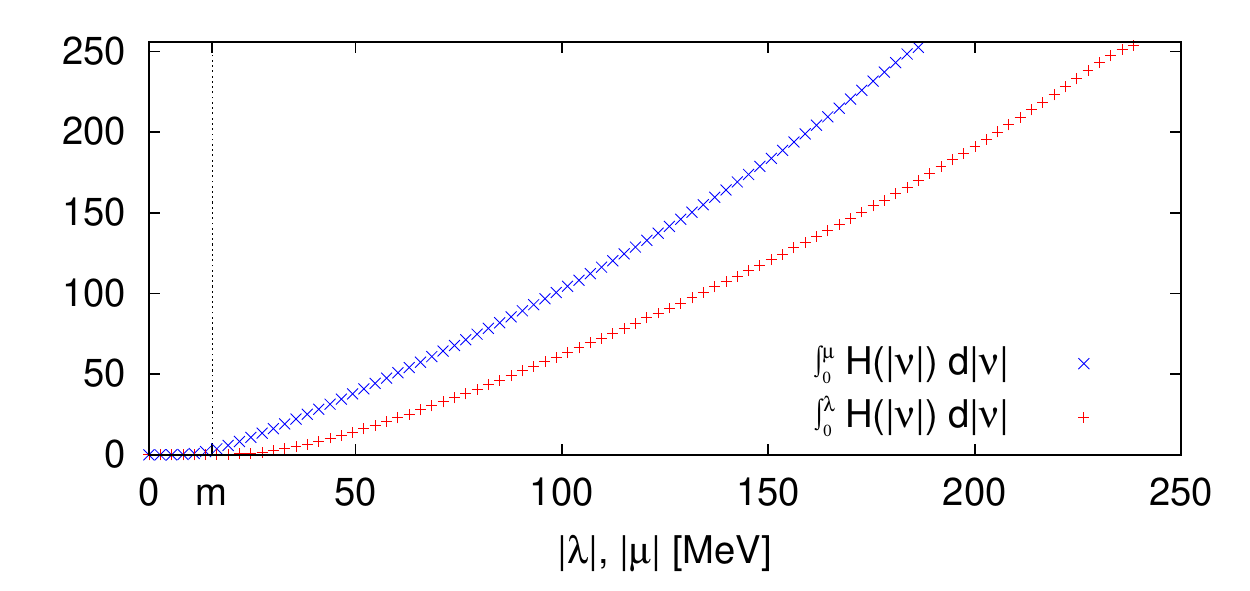}
\caption{L.h.s.: histograms of the lowest 256 eigenvalues of $D$ (red) and $D_5$ (blue). R.h.s.: integrated histograms.}\label{histo}
\end{figure}

In \fig{histo} we show histograms of the lowest 256 eigenvalues of $D$ and $D_5$. Additionally in the same figure the integrals over
these histograms are given.

\section{Results}
\begin{figure}[h]
\centering
\includegraphics[width=0.5\columnwidth]{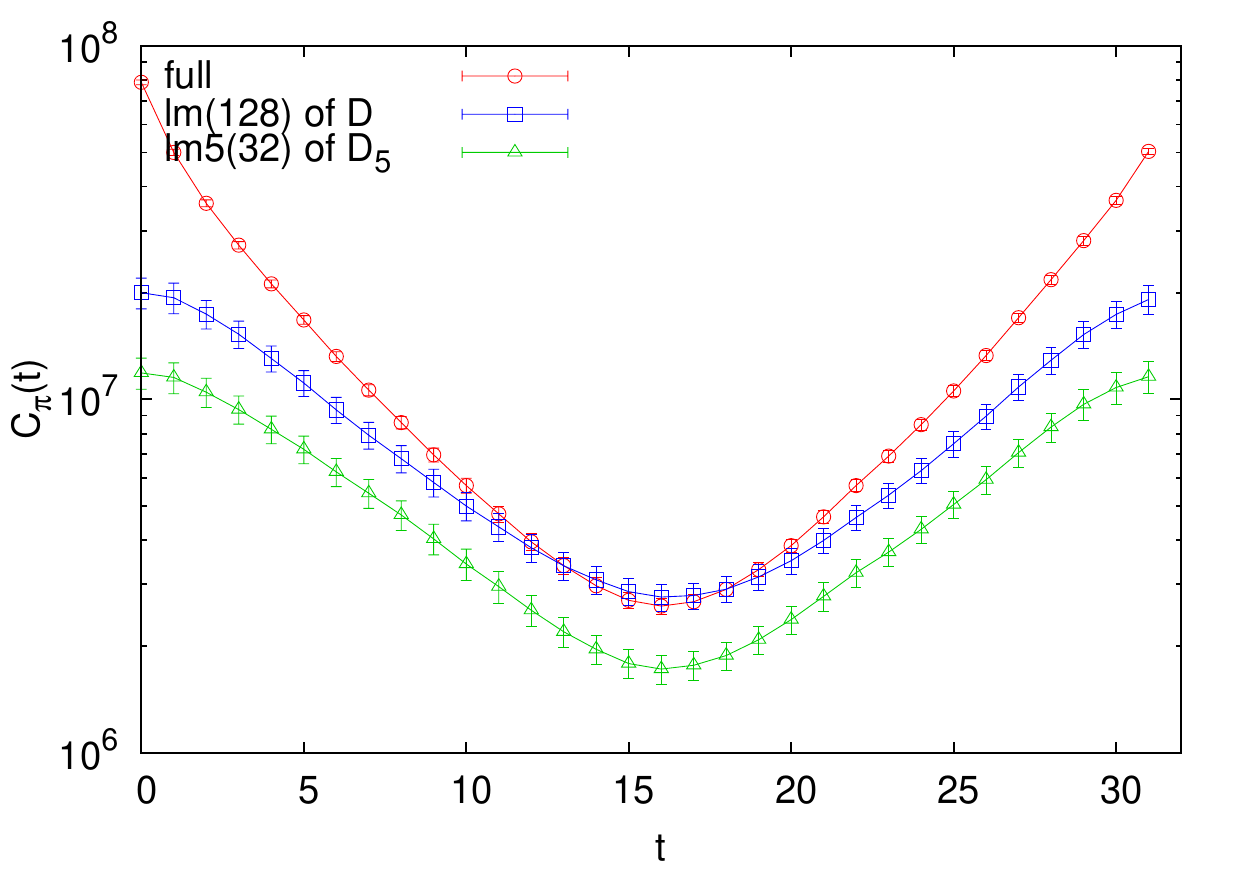}\hfill
\includegraphics[width=0.5\columnwidth]{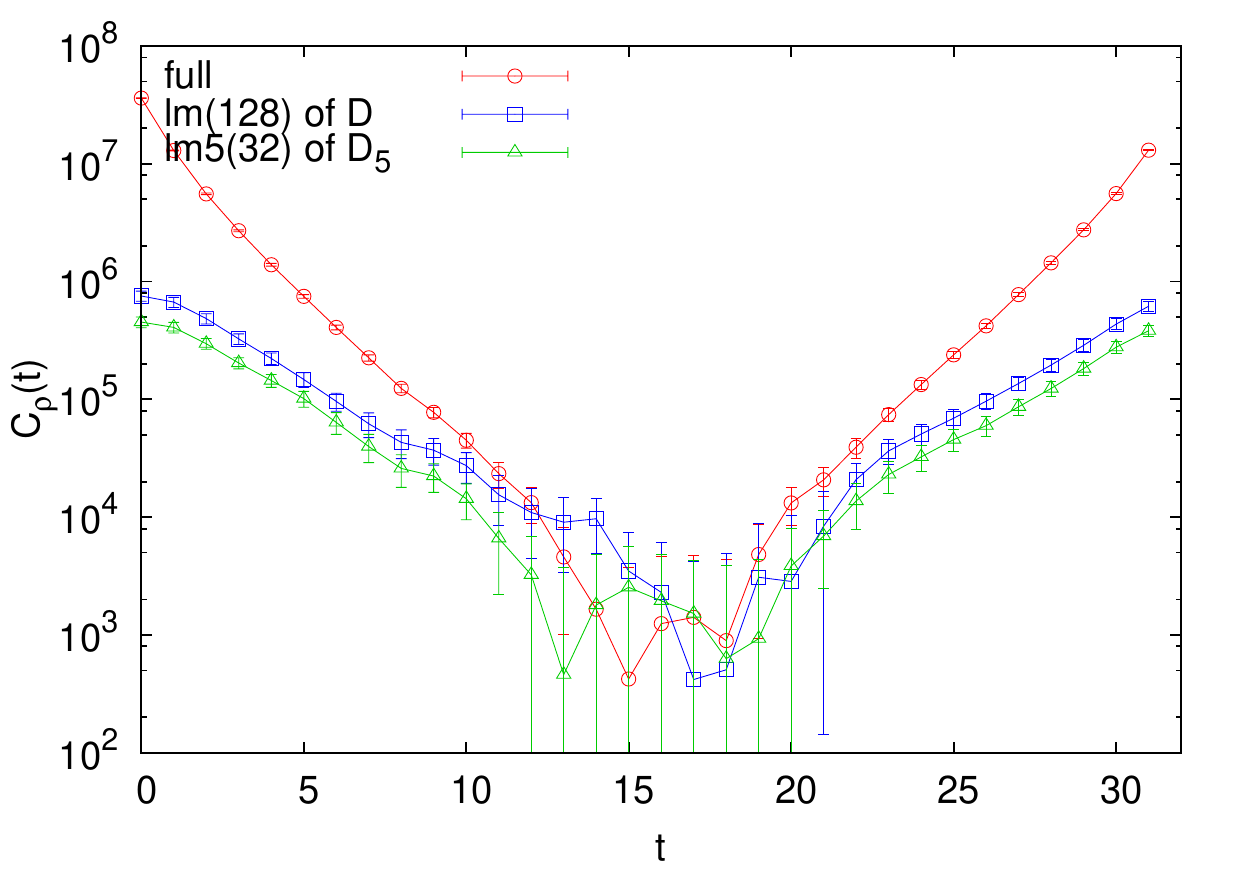}
\caption{Low mode contribution to the pion (left) and rho (right) correlators.}\label{LM}
\end{figure}

Before we study the effect of low mode reduction, we look at meson correlators which are build solely out of the low mode part
of the propagator, $S_\LM{k}$ or $S_\LMH{k}$, respectively (following the ideas in \cite{DeGrand:2000gq,DeGrand:2001tm,DeGrand:2003sf}).
In \fig{LM} we show the pion and rho correlators (interpolators one and three of Table\ref{tab:mesontab}) from
the low mode only propagators $S_\LM{128}$ and $S_\LMH{32}$ in comparison with the correlators from full propagators.
As can be seen in both cases, four times less eigenvalues of $D_5$ in 
comparison to eigenvalues of $D$ 
are already sufficient to obtain a similar quality approximation of the full correlators.
Since the eigenvalues of $D_5$ seem to be more effective for the low mode saturation of meson correlators,
we will concentrate on the eigenvalues of $D_5$ throughout the rest of this work.

\begin{figure}[h]
\centering
\includegraphics[width=0.5\columnwidth]{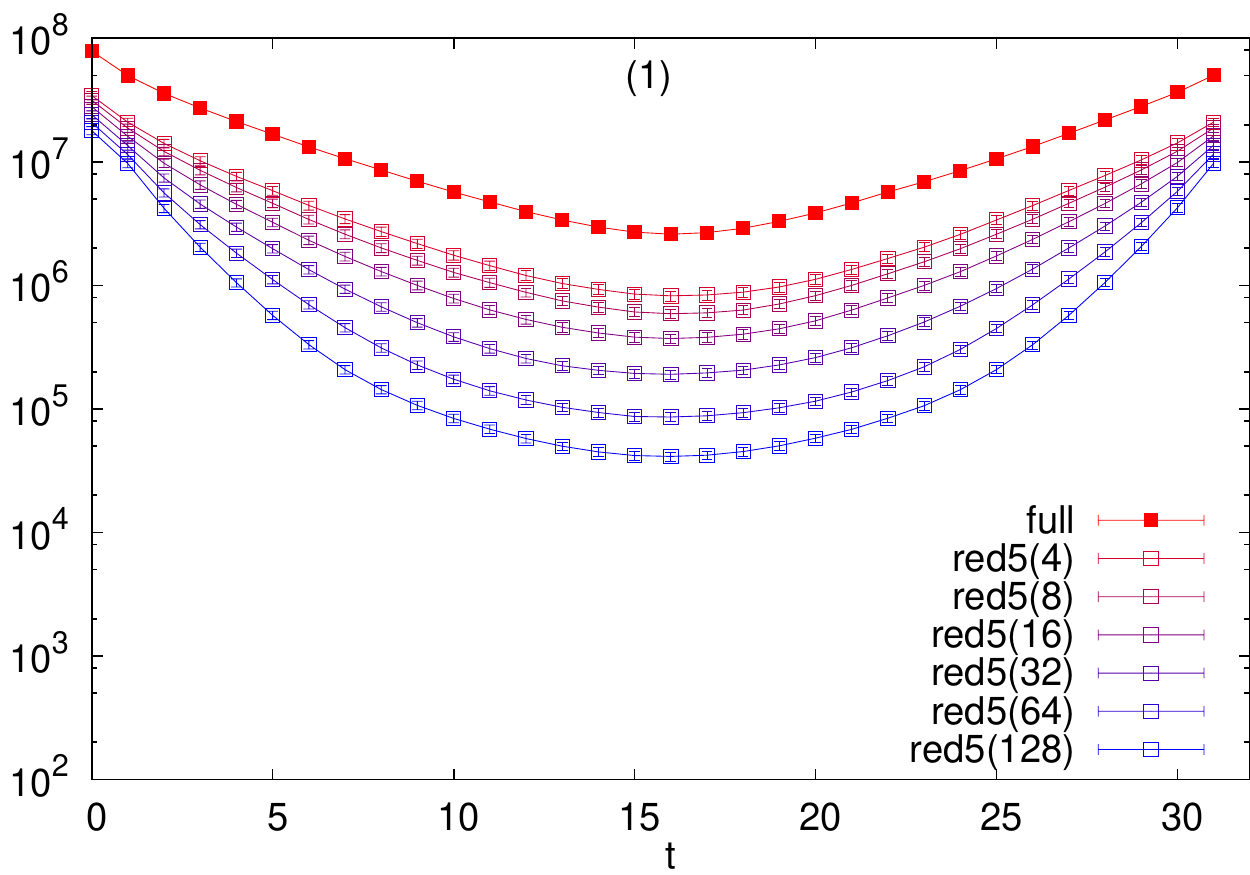}\hfill
\includegraphics[width=0.5\columnwidth]{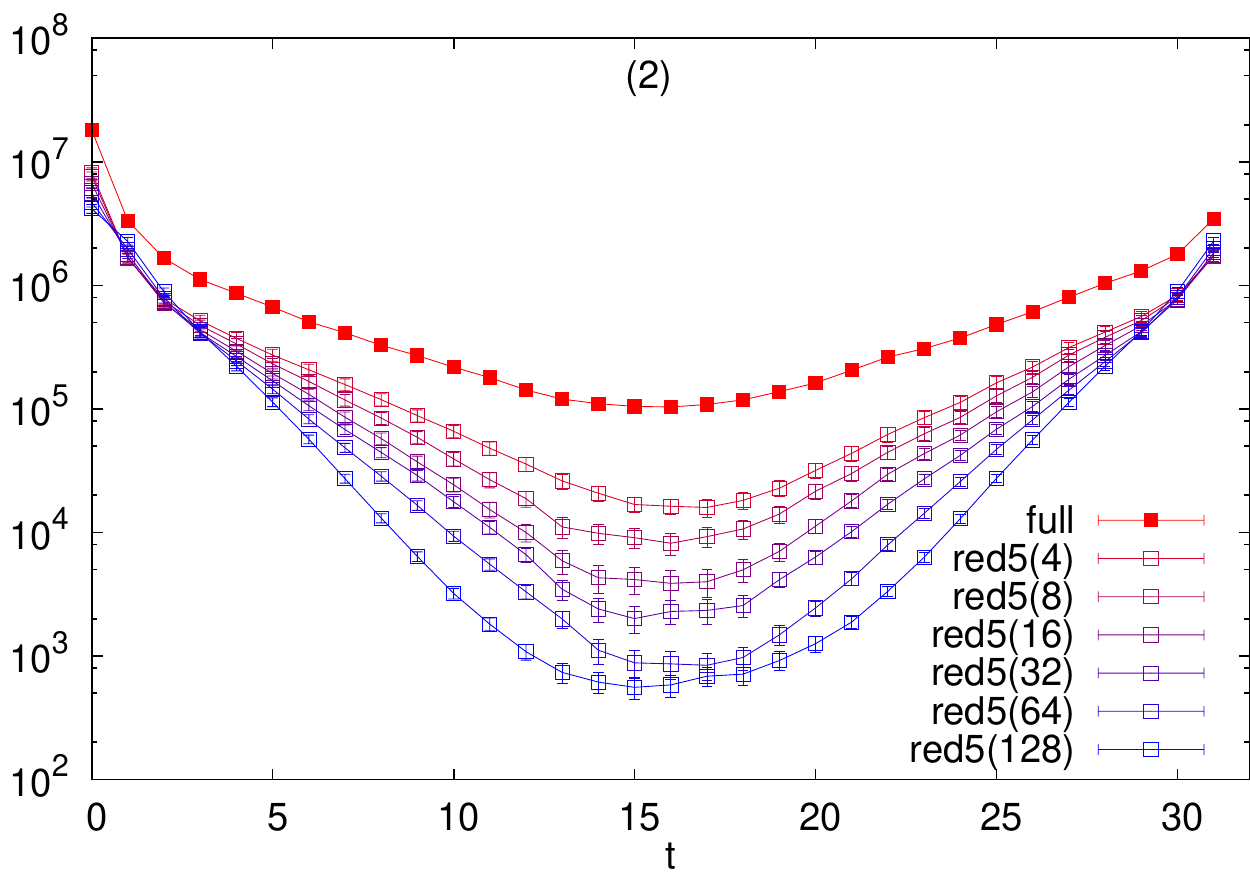}\\
\includegraphics[width=0.5\columnwidth]{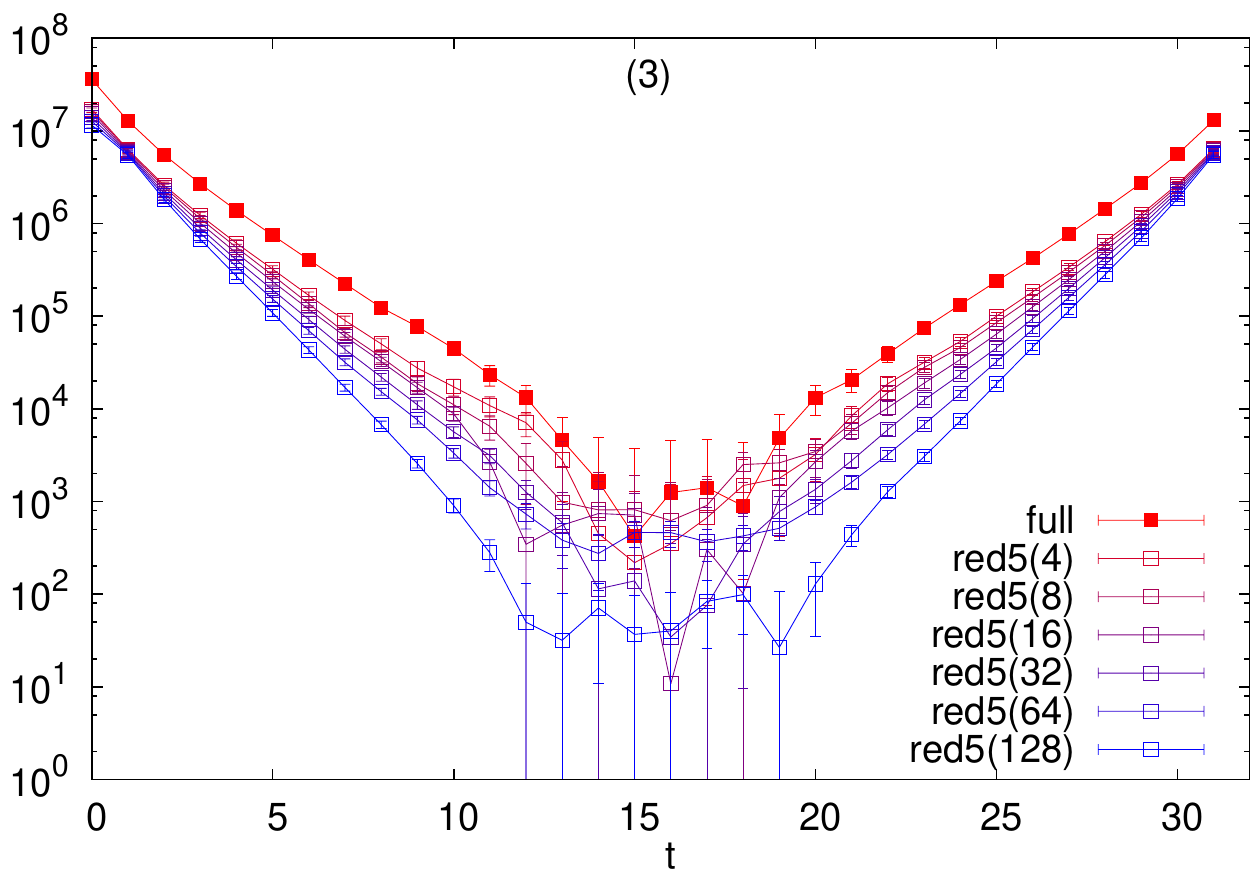}\hfill
\includegraphics[width=0.5\columnwidth]{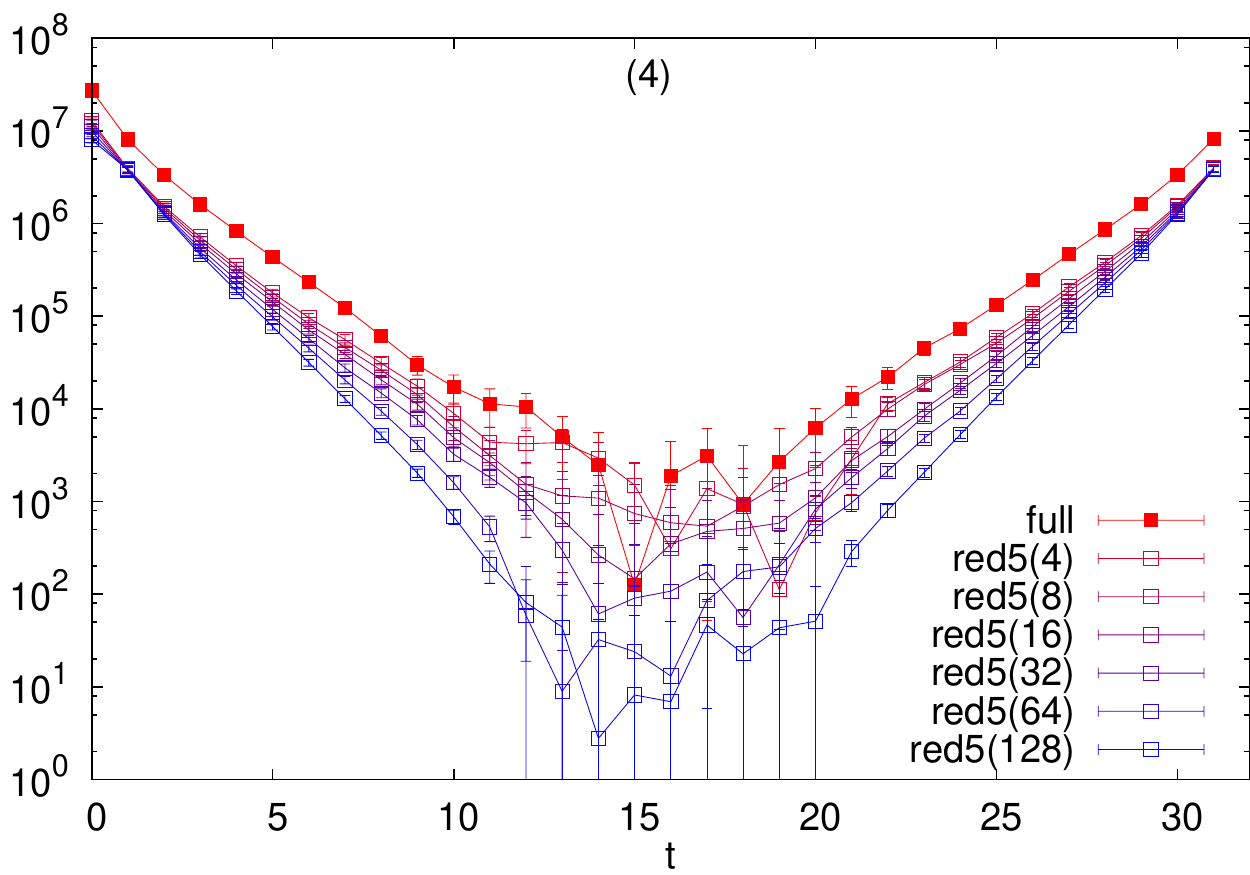}\\
\includegraphics[width=0.5\columnwidth]{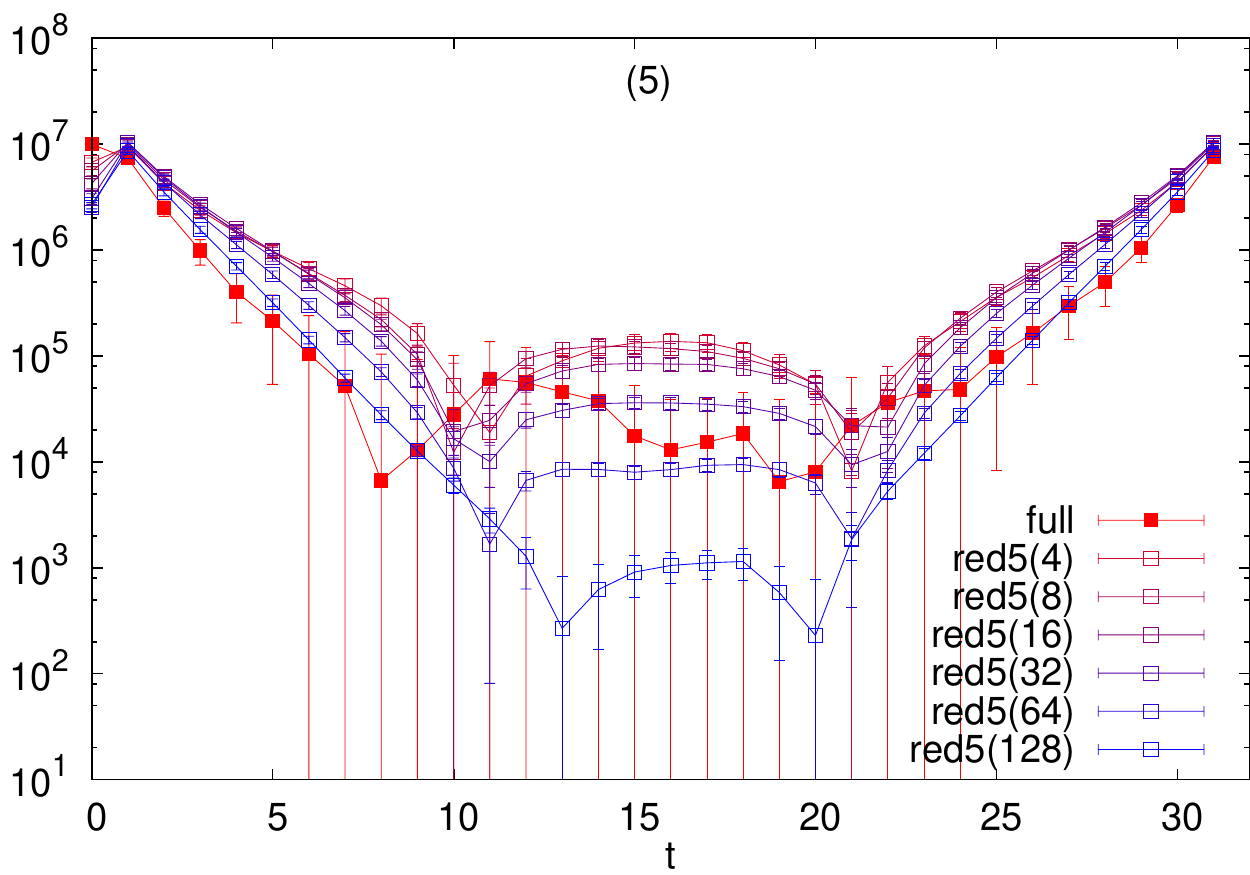}\hfill
\includegraphics[width=0.5\columnwidth]{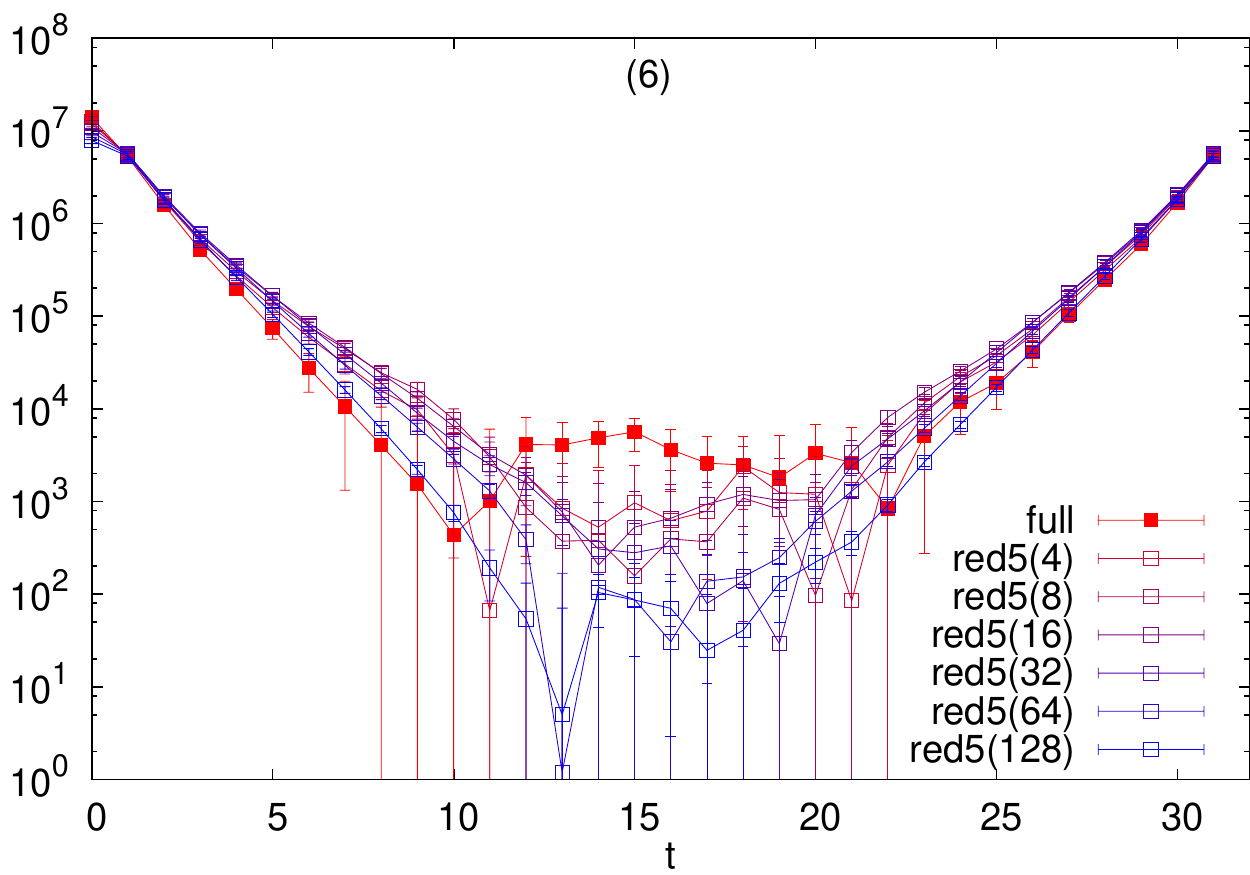}
\caption{Correlators under low mode reduction. The number in parentheses indicates which current is considered, 
cf. Table \protect\ref{tab:mesontab}.}\label{corr}
\end{figure}

In \fig{corr} we show the correlators under $D_5$ low mode \emph{reduction} in comparison to the correlators from full quark propagators.
We find that negative parity mesons (currents 1--4) become heavier under low mode reduction
whereas positive parity mesons (currents 5--6) at first become lighter and after subtracting more than 8--16 modes,
their mass begins to increase as well.
Except for the pion from current one, all correlators still show a clear exponential form (at least as good as for
the full propagators) from which we conclude
that we are still confronted with bound particles. 
The loss of the exponential behavior in current one can qualitatively be explained
with the strong sensitivity of the pion on the lowest Dirac modes in the sense that very few modes are sufficient
to saturate the pseudoscalar correlator in the medium to large time region.
Moreover we find that low mode removal reduces the noise of the correlators which is most obvious
for the positive parity currents.

\begin{wrapfigure}{r}{0.5\textwidth}
\centering
\includegraphics[width=0.5\columnwidth]{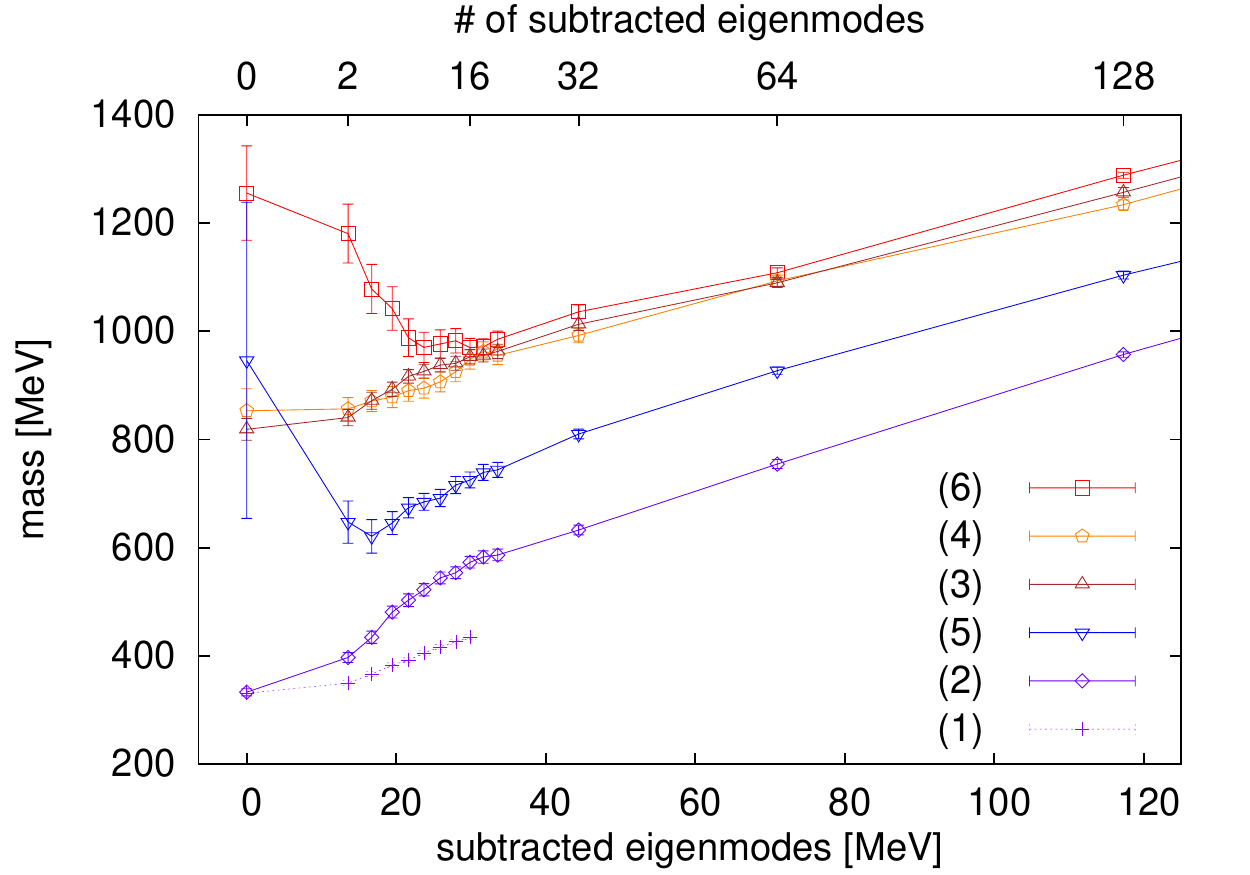}
\caption{The change of the masses of the mesons under low mode removal of the ingoing quark propagators.
The labeling number corresponds to the currents from Table \protect\ref{tab:mesontab}.}\label{masses}
\end{wrapfigure}
We performed effective mass fits for all the reduced correlators and combined the resulting mass values,
as a function of the reduction level, in \fig{masses}.
For the pion from current one we were only able to perform fits for very low reduction levels.
The masses of current three and six are found to be degenerate from truncation level 16 onwards
which
corresponds to having subtracted all eigenvalues with a magnitude equal to or less than twice the 
bare quark mass of our setting. Furthermore current four becomes degenerate with currents three and six
which hints towards a larger degeneracy which also might include the $h_1$ which we did not calculate.
Whereas we observe chiral restoration in the $J=1$ states, we cannot draw any conclusions for $J=0$ states since
the would-be chiral partner of the $a_0$ (current five) is the pion (current one) which gets destroyed under low mode reduction.
Current two which couples to the pion due to PCAC (partially conserved vector current) does not mix with the scalar
so there is no reason these two states should become degenerate. For further discussion see \cite{Lang:2011qy}.

\section{Conclusions}
We studied the isovector meson spectrum using valence quark propagators 
where  a varying number of the
lowest Dirac eigenmodes has been removed.
The density of the lowest eigenmodes is related to the spontaneous breaking of the chiral symmetry \cite{Banks:1979yr}.
We find restoration of the chiral symmetry in $J=1$ states when only 16 eigenmodes have been subtracted whereas
the exponential form of the correlators does not suffer from this modification. An exception is
the pion correlator for which the lowest Dirac eigenmodes are crucial, its exponential behavior rapidly gets lost when subtracting low modes. We conclude that the low lying Dirac modes -- although providing an important signal for spontaneous chiral symmetry breaking -- are not important for
the binding mechanism of quarks at least in the isovector mesons (with the exception of the pion).

\begin{acknowledgments}
We would like to thank L. Glozman for helpful discussions.  The
calculations have been performed on the SGI Altix 4700 of the
LRZ Munich and on clusters at ZID at the University of Graz. Support by
DFG SFB-TR55 and by Austrian Science Fund (FWF) DK W1203-N16
is gratefully acknowledged. M.S. is supported by the Research Executive 
Agency (REA) of the European Union under Grant Agreement 
PITN-GA-2009-238353 (ITN STRONGnet).

\end{acknowledgments}


\begin{thebibliography}{10}

\bibitem{Banks:1979yr}
T.~Banks and A.~Casher, {\it Chiral symmetry breaking in confining theories},
  {\em Nucl. Phys. B} {\bf 169} (1980) 103.

\bibitem{DeGrand:2000gq}
T.~A. DeGrand and A.~Hasenfratz, {\it {Low lying fermion modes, topology and
  light hadrons in quenched QCD}},  {\em Phys. Rev. D} {\bf 64} (2001) 034512,
  [\href{http://arxiv.org/abs/hep-lat/0012021}{{\tt hep-lat/0012021}}].

\bibitem{DeGrand:2001tm}
T.~DeGrand, {\it {Short distance current correlators: Comparing lattice
  simulations to the instanton liquid}},  {\em Phys. Rev. D} {\bf 64} (2001)
  094508, [\href{http://arxiv.org/abs/hep-lat/0106001}{{\tt hep-lat/0106001}}].

\bibitem{DeGrand:2003sf}
T.~A. DeGrand, {\it {Eigenvalue decomposition of meson correlators}},  {\em
  Phys. Rev. D} {\bf 69} (2004) 074024,
  [\href{http://arxiv.org/abs/hep-ph/0310303}{{\tt hep-ph/0310303}}].

\bibitem{DeGrand:2004qw}
T.~DeGrand and S.~Schaefer, {\it {I}mproving meson two-point functions in
  lattice {QCD}},  {\em Comput. Phys. Commun.} {\bf 159} (2004) 185--191,
  [\href{http://arxiv.org/abs/hep-lat/0401011}{{\tt hep-lat/0401011}}].

\bibitem{DeGrand:2004wh}
T.~DeGrand and S.~Schaefer, {\it {I}mproving meson two-point functions by
  low-mode averaging},  {\em Nucl. Phys. (Proc. Suppl.)} {\bf 140} (2005) 296,
  [\href{http://arxiv.org/abs/hep-lat/0409056}{{\tt hep-lat/0409056}}].

\bibitem{Giusti:2004yp}
L.~Giusti, P.~Hern\'andez, M.~Laine, P.~Weisz, and H.~Wittig, {\it {L}ow-energy
  couplings of {QCD} from current correlators near the chiral limit},  {\em
  JHEP} {\bf 04} (2004) 013, [\href{http://arxiv.org/abs/hep-lat/0402002}{{\tt
  hep-lat/0402002}}].

\bibitem{Neuberger:1997fp}
H.~Neuberger, {\it {E}xactly massless quarks on the lattice},  {\em Phys. Lett.
  B} {\bf 417} (1998) 141, [\href{http://arxiv.org/abs/hep-lat/9707022}{{\tt
  hep-lat/9707022}}].

\bibitem{Neuberger:1998wv}
H.~Neuberger, {\it {M}ore about exactly massless quarks on the lattice},  {\em
  Phys. Lett. B} {\bf 427} (1998) 353,
  [\href{http://arxiv.org/abs/hep-lat/9801031}{{\tt hep-lat/9801031}}].

\bibitem{Lang:2011qy}
C.~B. Lang and M.~Schr\"ock, {\it Unbreaking chiral symmetry},  {\em Phys. Rev.
  D} {\bf 84} (2011) 087704, [\href{http://arxiv.org/abs/1107.5195}{{\tt
  arXiv:1107.5195}}].

\bibitem{Glozman:2007ek}
L.~Y. Glozman, {\it {R}estoration of chiral and ${U}(1)_{A}$ symmetries in
  excited hadrons},  {\em Phys. Rep.} {\bf 444} (2007) 1,
  [\href{http://arxiv.org/abs/hep-ph/0701081}{{\tt hep-ph/0701081}}].

\bibitem{Gattringer:2008vj}
C.~Gattringer, C.~Hagen, C.~B. Lang, M.~Limmer, D.~Mohler, and A.~Sch{\"a}fer,
  {\it Hadron spectroscopy with dynamical chirally improved fermions},  {\em
  Phys. Rev. D} {\bf 79} (2009) 054501,
  [\href{http://arxiv.org/abs/0812.1681}{{\tt arXiv:0812.1681}}].

\bibitem{Engel:2010my}
G.~P. Engel, C.~B. Lang,
  M.~Limmer, D.~Mohler, and A.~Sch{\"a}fer, {\it {Meson and baryon spectrum for
  QCD with two light dynamical quarks}},  {\em Phys. Rev. D} {\bf 82} (2010)
  034505, [\href{http://arxiv.org/abs/1005.1748}{{\tt arXiv:1005.1748}}].

\bibitem{Gattringer:2000js}
C.~Gattringer, {\it {A} new approach to {G}insparg-{W}ilson fermions},  {\em
  Phys. Rev. D} {\bf 63} (2001) 114501,
  [\href{http://arxiv.org/abs/hep-lat/0003005}{{\tt hep-lat/0003005}}].

\bibitem{Gattringer:2000qu}
C.~Gattringer, I.~Hip, and C.~B. Lang, {\it {A}pproximate {G}insparg-{W}ilson
  fermions: {A} first test},  {\em Nucl. Phys. B} {\bf 597} (2001) 451,
  [\href{http://arxiv.org/abs/hep-lat/0007042}{{\tt hep-lat/0007042}}].

\bibitem{LeSoYa98}
R.~B. Lehoucq, D.~C. Sorensen, and C.~Yang, {\em {ARPACK} {U}sers' {G}uide:
  Solution of large-scale eigenvalue problems with implicitly restarted Arnoldi
  methods}.
\newblock SIAM, New York, 1998.

\end{thebibliography}
\providecommand{\href}[2]{#2}\begingroup\raggedright\endgroup

\end{document}